# Tuning carbon nanotube bandgaps with strain

E. D. Minot, Yuval Yaish, Vera Sazonova, Ji-Yong Park, Markus Brink, Paul L. McEuen

*Laboratory of Atomic and Solid-State Physics, Cornell University, Ithaca, New York 14853*

We show that the band structure of a carbon nanotube (NT) can be dramatically altered by mechanical strain. We employ an atomic force microscope tip to simultaneously vary the NT strain and to electrostatically gate the tube. We show that strain can open a bandgap in a metallic NT and modify the bandgap in a semiconducting NT. Theoretical work predicts that bandgap changes can range between ± 100 meV per 1% stretch, depending on NT chirality, and our measurements are consistent with this predicted range.

The electronic and mechanical properties of carbon NTs make them interesting for both technological applications and basic science. A NT can be either metallic or semiconducting depending on the orientation between the atomic lattice and the tube axis [1, 2]. NTs can accommodate very large mechanical strains [3] and have an extremely high Young's modulus [4]. Both theory and experiment indicate that NTs also have interesting electromechanical properties [5-12]. A pioneering experiment [10] showed that the conductance of a metallic NT could decrease by orders of magnitude when strained by an atomic force microscope (AFM) tip. The authors suggest that a local distortion of the $sp^2$ bonding where the NT is touched by the AFM tip causes the drop in conductance. In Ref. [12], however, it is argued that the observed drop in conductance is due to a bandgap induced in the NT as it is axially stretched [5, 8, 11] as illustrated in Fig. 1(a). Evidence for the effect of strain on NT bandgap also comes from recent STM work on semiconducting NTs containing encapsulated metallofullerenes [13]. The authors found a bandgap reduction of 60% at the expected positions of the metallofullerenes and postulated that strain could account for this change.

Here we present measurements to demonstrate conclusively that strain modulates the band structure of NTs. We employ an AFM tip to simultaneously vary the NT strain and to electrostatically gate the tube. We find that, under strain, the conductance of the NT can increase or decrease, depending on the tube. By using the tip as a gate, we show that this is related to the increase or decrease in the bandgap of a NT under strain. The magnitude of the effect and its dependence on strain are consistent with theoretical expectations.

The samples consist of NTs suspended over a trench and clamped at both ends by electrical contacts [10, 14-17]. CVD growth is utilized to grow NTs with diameters between 1 and 10 nm at lithographically defined catalyst sites [18] on a Si substrate with a 500nm oxide. Metal contacts (5nm Cr and 50-80nm gold) are made using photolithography, as described previously [19]. An ashing step (400°C for 10 minutes in Ar atmosphere) removes photoresist residue and improves contact resistances. An HF etch (3 minutes in 6:1 BHF, etch rate 80 nm/min) followed by critical point drying is used to suspend the NTs [16]. Device conductances are not changed significantly by the etching/drying procedure [20].

The suspended tubes are then placed in a commercial AFM system (Dimension 3100, Digital Instruments) for simultaneous electrical measurements and AFM imaging/manipulation [21]. The AFM tips used (Nanosensor EFM tip) have a nominal radius of 20 nm, cantilever spring constants of 1-5 N/m, and are coated with a PtIr metal layer. Spring constants were calibrated using the thermal noise method [22]. To probe the mechanical and electromechanical properties, the AFM tip is centered above a suspended NT using a tapping mode image for guidance. The tip is then moved in the z-direction as illustrated in Fig. 1(b), while monitoring the static deflection of the cantilever and/or the conductance of the tube [23].

We first discuss the mechanical response of the NTs. Plotted in Fig. 2 is the upward force on the cantilever $F_z$, as a function of the tip height $z$, while raising the AFM tip. $F_z$ and $z$ are determined from measurements of the tip deflection, the cantilever spring constant, and the z-piezo motion. Figure 2 shows that the force the tube exerts on the tip can be both positive (upward) as well as negative (downward). These are separated by a region of near-zero force when the tip is near the plane of the contacts.

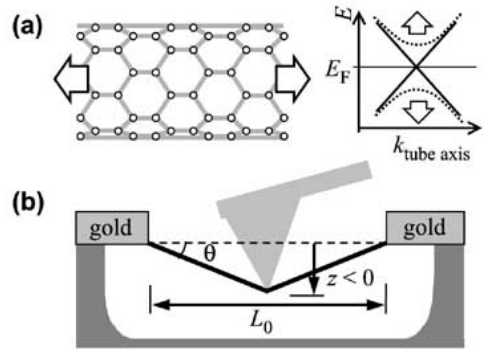

FIG. 1. (a) Real space representation of a zig-zag NT and the dispersion relation near $E_F$. Solid lines and dashed lines show $E(k)$ before and after stretching, respectively. (b) Experimental geometry for applying strain and gate voltage with an AFM tip, and measuring conductance with gold contacts. $L_0$ is distance between anchoring points, $z$ is the distance the center of the NT is displaced from the plane of the anchoring points.

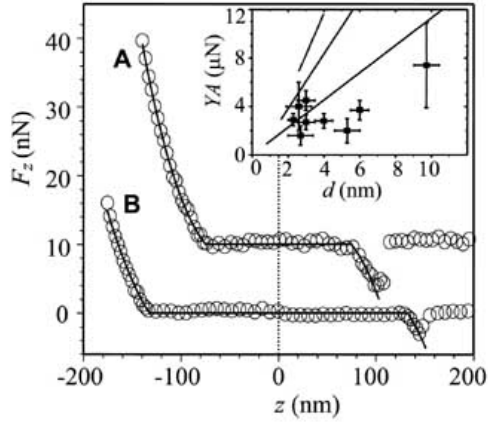

FIG. 2. Force-distance curves of two NTs; curve A is offset for clarity. Open circles show the measured deflection force as the AFM tip retracts towards positive $z$. Curve A is from a NT with $d = 5.3 \pm 0.5$ nm, $L_0 = 1.0 \pm 0.1$ μm. Curve B is from a NT with $d = 2.3 \pm 0.5$ nm, $L_0 = 1.5 \pm 0.1$ μm. Solid lines are fit curves given by Eq. 2. In curve A, slack = 11nm, $YA = 2$ μN and curve B slack = 22 nm and $YA = 2.9$ μN. The tubes were not measured by the same AFM cantilever. The inset shows a summary of all measured strain-tension proportionality constants, $YA$. Solid lines in the inset show predicted $YA$ values for single-, double- and triple-walled NTs of outer diameter $d$ when all shells carry the same mechanical load.

Force-distance curves such as Fig. 2 indicate that there is significant adhesion between the AFM tip and the NT. This adhesion, likely due to van der Waals forces, has been observed and discussed by other authors [24]. As the tip is raised above $z = 0$ the tube pulls downward on the tip before a sudden release. The adhesion serves an important function; from the distance between pushing and pulling onsets, $\pm z_{onset}$, the "slack" of a suspended NT can be determined. The slack is defined as $L_{tube} - L_0$ where $L_{tube}$, the tube length, is greater than $L_0$, the separation between the anchoring points. Nearly all NTs measured were slack, with typically 5-10 nm of slack for a 1μm tube. The slack is consistant with the slightly curved paths NTs followed across the oxide surface before etching. From $L_{tube}$ and $L_0$, axial strain $\sigma(z)$ is readily calculated:

$$\sigma(z) = \left(\sqrt{L_0^2 + 4z^2} - L_{tube}\right)/L_{tube} \quad (1)$$

for $|z| > z_{onset}$.

We find that force-distance curves such as Fig. 2 could be accurately fit by ignoring the bending modulus of the tube and assuming a linear proportionality between NT tension $T$ and axial strain $\sigma$. We write the proportionality constant as $YA$, where $Y$ is an effective Young's modulus, and $A$ is an effective cross-sectional area. The fitted curve for upward deflection force is given by:

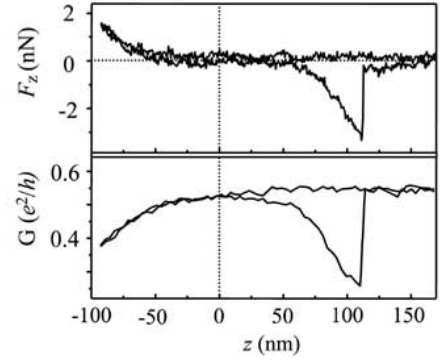

FIG. 3. Deflection force and conductance as a function of tip height for a NT with $d = 6.5$ nm and $L_0 = 1.9$ μm at $V_{tip} = 0$ V. The tip was first moved toward the surface (toward negative $z$) and then away from the surface.

$$F_z(z) = 2T\sin\theta = 2(YA \cdot \sigma(z))\left(\frac{2z}{\sqrt{L_0^2 + 4z^2}}\right) \text{ for } |z| \geq z_{onset}$$

$$F_z(z) = 0 \quad \text{for } |z| \leq z_{onset} \quad (2)$$

Fits are plotted as solid lines in Fig. 2; the model has fit well to all NTs we have measured.

The relationship between the fitting parameter $YA$ and NT diameter $d$ is shown in the inset of Fig. 2 [25]. The inset to Fig. 2 also contains the predicted values for $YA$ for NTs with one, two and three walls contributing to the tension [26-28].

The magnitude of $YA$ values and linearity with diameter $d$ indicate that a single shell is most likely carrying the mechanical load. This is true even for large diameter tubes, which likely have multiple walls. Similar observations of $YA$ magnitude [4] and outer-shell loading [3] have been reported.

We now turn to the electromechanical response of the NTs. The experiments are the same as those described above, but with the addition of in-situ conductance monitoring and control of AFM tip voltage $V_{tip}$. A lock-in amplifier operating with 10mV bias at 1kHz is employed to monitor NT conductance $G$. Figure 3 shows both $F_z$ and $G$, while engaging and then retracting from a NT. $V_{tip}$ is held at 0 V. When the cantilever is deflected, the $G$ is lowered, in agreement with previous results [10]. Interestingly, other tubes showed different behavior. Of the seven samples studied, two semiconducting tubes showed *increasing* $G$ when strained, one semiconducting and two metallic tubes showed decreasing $G$ when strained, and two metallic tubes showed no significant change.

To understand the origin of this behavior, the tip is used as a gate to investigate the band structure of the NT under strain. $V_{tip}$ is swept $\sim 3$ times per second over a range of a few volts as strain is slowly increased [29]. $G$ vs. $V_{tip}$ for different strains are shown in Fig. 4 for two NTs. At no strain ($\sigma = 0$) the observed $V_{tip}$ dependence indicates that the tube of Fig. 4(a) is metallic ($G$ is independent of $V_{tip}$) and the tube of Fig. 4(b) is semiconducting ($G$ is strongly dependent on $V_{tip}$). As the metallic tube is strained, an asymmetric dip centered at $V_{tip} \approx 1$V develops in $G$-$V_{tip}$. On the other hand, the semiconducting tube, which has an asymmetric minimum at $V_{tip} \approx 1$V without strain, shows an increase of $G$ with strain and a reduction in the asymmetry of the dip. The insets show the maximum resistance $R_{max}(\sigma)$ for each sweep of $V_{tip}$ as a function of strain, along with a fit to the functional form:

$$R_{max}(\sigma) = R_0 + R_1 \exp(\beta\sigma) . \quad (3)$$

This functional form fits the data well in both cases, but with different values of $R_0$, $R_1$ and $\beta$.

These results can be understood by the effect of strain on the bandgap of the tube, as described by previous authors [5, 8, 11]. These papers predict a chirality dependent value for the rate of change of bandgap with respect to strain, $dE_{gap}/d\sigma$. From Ref. [11]:

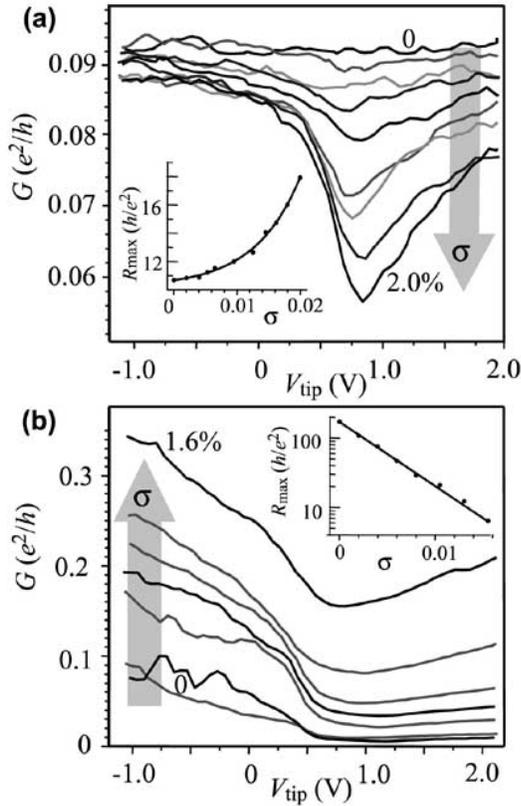

$$\frac{dE_{gap}}{d\sigma} = \text{sign}(2p+1)3t_0(1+\upsilon)\cos 3\phi , \quad (4)$$

where $t_0 \approx 2.7$eV is the tight-binding overlap integral, $\upsilon \approx 0.2$ is the Poisson ratio, $\phi$ is NT chiral angle and $p = -1$, 0 or 1 such that the wrapping indices, $n_1$ and $n_2$ satisfy $n_1 - n_2 = 3q + p$ where $q$ is integer [11]. The maximum value of $|dE_{gap}/d\sigma|$ is $3t_0(1+\upsilon) \approx 100$ meV/%, similar in magnitude to $|dE_{gap}/d\sigma|$ of typical bulk semiconductors. Note that half of all semiconducting NTs ($p = 1$) will have $dE_{gap}/d\sigma > 0$, while the other half ($p = -1$) have $dE_{gap}/d\sigma < 0$.

With knowledge of this electromechanical effect we first interpret the constant-tip-voltage experiments. When strain causes the measured conductance to drop, $dE_{gap}/d\sigma$ is positive. If conductance increases, $dE_{gap}/d\sigma$ is negative. Lastly, if conductance does not change with strain, then $dE_{gap}/d\sigma$ is zero or close enough to zero to be undetectable.

The $G$-$V_{tip}$ curves shown in Fig. 4 confirm this picture. The curves can be qualitatively understood by combining the strain-induced bandgap modification with standard concepts in semiconducting NT transport. We assume that the tip gate affects the middle portion of tube while sections close to the contacts are held p-type [30]. Sweeping $V_{tip}$ will change the strained tube from a p-doped semiconductor ($V_{tip} = 0$ V, Fig. 5(a)) to a p-n-p junction ($V_{tip} > 1$ V, Fig. 5(c)). The conductance minimum at $\sim 1$ V corresponds to charge carrier depletion in the NT's middle section during the transition from p-type to n-type (Fig. 5(b)); the size of the conductance dip depends on the bandgap $E_{gap}$, which changes with strain.

A model of the resistance $R_{max}(\sigma)$ associated with this conductance minimum can be used to determine $dE_{gap}/d\sigma$ and other transport parameters of the tube. We first consider transport by thermal activation, neglecting tunneling across the depleted region. Referring to Fig. 5(b), electrons with energy $E$ such that $|E - E_F| > E_{gap}$ cross the barrier with a transmission probability $|t|^2$, while those with $|E - E_F| < E_{gap}$, have zero transmission probability. The low-bias resistance of the device is then:

$$R_{tot} = R_S + \frac{1}{|t|^2} \frac{h}{8e^2}\left(1 + \exp\left(\frac{E_{gap}}{kT}\right)\right), \quad (5)$$

where the first term $R_S$ is the resistance in series with the junction due to the metal-NT contacts, etc., and the second term is the resistance of the junction region. From Eq. 4 we have:

$$E_{gap} = E_{gap}^0 + \frac{dE_{gap}}{d\sigma}\sigma. \quad (6)$$

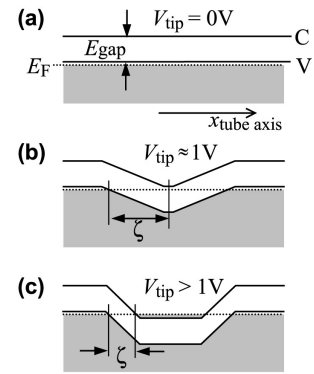

FIG. 4. (a) $G$-$V_{tip}$ measurements of a NT with $d = 3 \pm 0.5$ nm and $L_0 = 1.4 \pm 0.1$ μm at 0, 0.5, 0.7, 1.0, 1.2, 1.4, 1.6, 1.8 and 2.0% strain. (b) $G$-$V_{tip}$ measurements of a NT with $d = 4 \pm 0.5$ nm and $L_0 = 1.1 \pm 0.1$ μm at 0, 0.2, 0.6, 0.8, 1.1, 1.3, and 1.6% strain. Closed circles in the insets show maximum device resistance as a function of strain. Solid lines in the insets show fit curves given by Eq. 3. For (a) $R_0 = 10.2h/e^2$, $R_1 = 0.5h/e^2$ and $\beta = 139$. For (b) $R_0 = 0$, $R_1 = 171h/e^2$ and $\beta = -213$.

FIG. 5. Evolution of the energy band diagram as $V_{tip}$ is increased. The tube is held p-type at the contacts. The valence and conduction band edges are denoted by V and C. (a) $V_{tip} = 0$ V, the NT is a p-type semiconductor. (b) $V_{tip} \approx 1$ V, transport is interrupted by a depleted region. (c) $V_{tip} > 1$ V, a p-n-p junction forms in the middle of the tube. Transport due to tunneling increases as the distance $\zeta$ becomes smaller.

Eq.'s (5) and (6) give physical meaning to the fitting parameters $\beta$, $R_0$ and $R_1$ that were introduced in Eq. (3). Most important is $\beta$, the exponential fitting parameter, which is related to the strain-dependence of the gap: $dE_{gap}/d\sigma = \beta kT$. From the measured values of $\beta$, we obtain $dE_{gap}/d\sigma = -53$ meV/% for the semiconducting tube in Fig. 4(b) and $dE_{gap}/d\sigma = +35$ meV/% for the metallic tube in Fig. 4(a). From Eq. (4) we can then estimate chiral angles $\phi \approx 19°$ and $23°$ for the two tubes respectively. For an accurate determination of the chiral angles additional experiments are necessary to verify the quantitative validity of Eqs. (4) and (5).

Additional knowledge about the device can be gained from the fitting parameter $R_1 = (h/8e^2|t|^2)\exp(E_{gap}^0/kT)$. Fitting results for the metallic tube ($E_{gap}^0 = 0$) gives $R_1 = 0.49h/e^2$, and hence a transmission probability $|t|^2 = 0.25$. Transport of thermally activated electrons across the junction region is thus not ballistic, but nevertheless highly transmissive, as expected from previous measurements of long mean free paths in NTs [31]. Fitting the semiconducting tube data yields a much higher resistance, $R_1 = 171h/e^2$. Using an estimate of $|t|^2 = 0.25$ from above, we infer $E_{gap}^0 = 160$ meV. This inferred energy gap corresponds to a tube with diameter $d = 4.7$nm (using $E_{gap}^0 = 2t_0 r_0/d$ [32]), in reasonable agreement with $d = 4 \pm 0.5$ nm measured by AFM. The agreement provides support for the validity of Eq. (5). However, variable temperature studies are needed to definitively separate out the tunneling and thermal activation contributions [33]; unfortunately, this is not possible with our current AFM.

These results demonstrate that strain can be used to continuously tune the bandgap of a NT. The ability to control bandgap should help to clarify our understanding of transport in NTs and also enable a number of applications. With further testing of Eqs. (4) and (5) measurements of $dR_{max}/d\sigma$ may be used to uniquely determine the wrapping indices of small-diameter NTs. Electrical transduction of small forces is also possible; for example, the most sensitive device studied here has a sensitivity of 0.1 nN/(Hz)$^{1/2}$ at low frequencies. Finally, NT heterostructures, where different sections of a single NT have different bandgaps, can be created if the different sections can be selectively strained. NTs are ideal for such applications because they can accommodate very large strains.

We wish to acknowledge useful discussions with Hande Ustunel, Tomas Arias, and Hongjie Dai. This work was supported by the NSF through the Cornell Center for Materials Research and the NSF Center for Nanoscale Systems, and by the MARCO Focused Research Center on Materials, Structures, and Devices. Sample fabrication was performed at the Cornell node of the National Nanofabrication Users Network, funded by NSF. One of us (E.D.M.) acknowledges support by an NSF Graduate Fellowship.

[1] T. W. Odom, J. L. Huang, P. Kim, et al., Nature **391**, 62 (1998).
[2] J. W. G. Wildoer, L. C. Venema, A. G. Rinzler, et al., Nature **391**, 59 (1998).
[3] M. F. Yu, O. Lourie, M. J. Dyer, et al., Science **287**, 637 (2000).
[4] For a review see J. P. Salvetat, J. M. Bonard, N. H. Thomson, et al., Appl. Phys. A **69**, 255 (1999).
[5] R. Heyd, A. Charlier, and E. McRae, Phys. Rev. B **55**, 6820 (1997).
[6] C. L. Kane and E. J. Mele, Phys. Rev. Lett. **78**, 1932 (1997).
[7] A. Rochefort, D. R. Salahub, and P. Avouris, Chem. Phys. Lett. **297**, 45 (1998).
[8] L. Yang, M. P. Anantram, J. Han, et al., Phys. Rev. B **60**, 13874 (1999).
[9] S. Paulson, M. R. Falvo, N. Snider, et al., Appl. Phys. Lett. **75**, 2936 (1999).
[10] T. W. Tombler, C. W. Zhou, L. Alexseyev, et al., Nature **405**, 769 (2000).
[11] L. Yang and J. Han, Phys. Rev. Lett. **85**, 154 (2000).
[12] A. Maiti, A. Svizhenko, and M. P. Anantram, Phys. Rev. Lett. **88**, 126805 (2002).
[13] J. Lee, H. Kim, S. J. Kahng, et al., Nature **415**, 1005 (2002).
[14] D. A. Walters, L. M. Ericson, M. J. Casavant, et al., Appl. Phys. Lett. **74**, 3803 (1999).
[15] J. P. Salvetat, G. A. D. Briggs, J. M. Bonard, et al., Phys. Rev. Lett. **82**, 944 (1999).
[16] J. Nygard and D. H. Cobden, Appl. Phys. Lett. **79**, 4216 (2001).
[17] N. R. Franklin, Q. Wang, T. W. Tombler, et al., Appl. Phys. Lett. **81**, 913 (2002).
[18] J. Kong, H. T. Soh, A. M. Cassell, et al., Nature **395**, 878 (1998).
[19] S. Rosenblatt, Y. Yaish, J. Park, et al., Nano Lett. **2**, 869 (2002).
[20] Many of the etched devices are not robust, as evidenced by degradation of electrical contact during AFM imaging. The data presented here is from NTs that have robust contact with the gold electrodes.
[21] J. Y. Park, Y. Yaish, M. Brink, et al., Appl. Phys. Lett. **80**, 4446 (2002).
[22] J. L. Hutter and J. Bechhoefer, Rev. Sci. Instrum. **64**, 1868 (1993).
[23] For consistent NT-tip alignment the x-y piezo must be stabilized for many minutes. The cantilever oscillation signal is a useful compliment to the tip deflection signal when first pushing a NT. As a NT is pushed, tip deflection can jump suddenly toward zero. Slack is not changed by these events;

therefore we associate such jumps with slips of the NT-tip contact. Favorable contact is achieved by trial-and-error. A low noise laser, available by special request from Digital Instruments technical support, suppresses laser interference effects that can contaminate tip deflection data.

[24] S. Decossas, G. Cappello, G. Poignant, et al., Europhys. Lett. **53**, 742 (2001).
[25] Uncertainty in $AY$ comes from uncertainty in $L_0$, x-y tip positioning and the cantilever spring constant.
[26] B. I. Yakobson, C. J. Brabec, and J. Bernholc, Phys. Rev. Lett. **76**, 2511 (1996).
[27] X. Zhou, J. J. Zhou, and Z. C. Ou-Yang, Phys. Rev. B **62**, 13692 (2000).
[28] J. P. Lu, Phys. Rev. Lett. **79**, 1297 (1997).
[29] The PtIr coating on the AFM tip makes very poor electrical contact with the NT with typical leakage resistances > 10 GΩ.
[30] J. W. Park and P. L. McEuen, Appl. Phys. Lett. **79**, 1363 (2001).
[31] A. Bachtold, M. S. Fuhrer, S. Plyasunov, et al., Phys. Rev. Lett. **84**, 6082 (2000).
[32] C. T. White and J. W. Mintmire, Nature **394**, 29 (1998).
[33] We have investigated the assumption that electron tunneling does not contribute significantly to $R_{max}$ (Eq. (5), Fig. 5(b)) by calculating the tunneling resistance for different junction electric fields $\varepsilon = E_{gap}/\zeta$. For $\varepsilon < 10^7$ V/m (corresponding to $\zeta >$ 20nm when $E_{gap} =$ 200meV) the effect of tunneling is negligible. Similar calculations and estimates of $\zeta$ are found in F. Leonard and J. Tersoff, Phys. Rev. Lett. **83**, 5174 (1999) and A. A. Odintsov, Phys. Rev. Lett. **85**, 150 (2000).